\documentstyle[mprocl]{article}

\def\Journal#1#2#3#4{{#1} {\bf #2}, #3 (#4)}

\def\PRL{\em Phys. Rev. Lett.}
\def\PRD{{\em Phys. Rev.} D}

\def\be{\begin{equation}}
\def\ee{\end{equation}}
\def\bea{\begin{eqnarray}}
\def\eea{\end{eqnarray}}

\begin{document}
\title{SELF-CONSISTENT SOLUTIONS FOR LOW-FREQUENCY GRAVITATIONAL 
BACKGROUND RADIATION}

\author{G. DAUTCOURT}

\address{Max-Planck-Institut f\"ur Gravitationsphysik 
(Albert-Einstein-Institut)\\
Schlaatzweg 1, 14473 Potsdam, Germany}


\maketitle\abstracts
{We study in a Brill-Hartle type of approximation the back-reaction
of a superposition of linear gravitational waves in an Einstein-de~Sitter
background up to the second order in the small wave amplitudes $h_{ik}$. 
The wave amplitudes are assumed to form a homogeneous and isotropic 
stochastic process. No restriction for the wavelengths is assumed. 
The effective stress-energy tensor $T^{e}_{\mu\nu}$ is
calculated in terms of the correlation functions of the process. 
We discuss in particular a situation where $T^{e}_{\mu\nu}$ is
the dominant excitation of the background metric. Apart from the Tolman 
radiation universe, a solution with the scale factor of the de~Sitter 
universe exists with $p = -\rho$ as effective equation of state.} 

While the description of cosmic gravitational radiation in linearized 
relativity is fairly well known, a study of its nonlinear aspects is
much harder. Numerical relativity\cite{an} is one way, but there are also
some analytical and semianalytical approaches. Already in 1964 Brill and
Hartle\cite{bh} proposed a scheme which takes the back-reaction of linear
gravitational waves into account. As it stands, the Brill-Hartle 
method is considered as a high-frequency approximation for gravitational
radiation. Thus its application to early inflationary stages of the
Universe is questionable, since low-frequency modes may be present, 
which turn into high-frequency modes (with wavelengths smaller than the 
temporary horizon) only at later time. 
As shortly discussed in this note, a slight modification of the 
Brill-Hartle approach can remove this shortcoming. The detailed 
calculations will be published elsewhere. 
The wave perturbations are considered as random variables, forming a 
stochastic process, which shares the symmetries of the background metric. 
The method is similar to the Monin-Yaglom\cite{my} approach to statistical 
fluid dynamics, and is also used in optical coherence theory\cite{mw}. 
For simplicity, an Einstein-de~Sitter model is chosen as background metric.
Tensor perturbations are added in synchronous gauge, assuming 
$g_{00}= -a^2$, $g_{0i}=0$, $g_{ik}= a^2\delta_{ik}+h_{ik},$ as well as
the gauge conditions $h_{ii}=0, h_{ik,k}=0$. $a(\eta)$ is the scale factor.   
The Einstein tensor may be expanded in powers of $h_{ik}$. Retaining 
terms up to second order and performing a stochastic (ensemble) average
on the field equations, they split into linear wave equations for 
the $h_{ik}$ and the back-reaction equations. The back-reaction 
equations relate the Einstein tensor for the background metric to the 
effective stress-energy tensor of the
waves, which is represented as stochastic average over bilinear
expressions in $h_{ik}$ and derivatives of $h_{ik}$. 
As well known, the effective stress-energy tensor is not gauge invariant 
in general \cite{}. However, as shown by Abramo, Brandenberger
and Mukhanov\cite{abm} (see also \cite{ab}, \cite{br}), 
the gauges change the background geometry to second order, and these 
changes just compensate the change of the stress-energy tensor. 
Representing the $h_{ik}$ as stochastic Fourier integrals 
$\int \gamma_{ik}({\bf k},\eta)e^{i{\bf kx}}d{\bf k} + conj.compl. $, 
the amplitudes $\gamma_{ik}$ satisfy an ordinary differential equation. 
The symmetry properties of the problem allow a simple representation of 
expressions which are bilinear in $h_{ik}$, e.g. 
\be
\langle h_{ik}h_{lm}\rangle
= \frac{16\pi a^2}{15}(3\delta_{il}\delta_{km}
+ 3 \delta_{im}\delta_{kl}-2\delta_{ik}\delta_{lm})
\int dk k^2 f, \label{eq:famp1} \nonumber \\
\ee 
in terms of a single spectral density $f(k,\eta)$. $f$ satisfies a
nonlinear differential equation ($\epsilon_0$ depends only on $k=|{\bf k}|$,
and a prime denotes the differentiation with respect to $\eta$)
\be 2ff'' - f'^2 +4f^2(k^2-\frac{a''}{a})-4\epsilon_0
= 0. \label{eq:al1}
\ee
For high-frequency waves $k\eta \gg 1$, $f=\sqrt{\epsilon_0}/k$ follows
as solution ("high-frequency approximation"). It is convenient to write 
$\rho$ and $p$ in the effective stress-energy tensor 
in terms of four frequency-independent, but in general time-dependent 
integrals ("moments") over the spectral density $f(k,\eta)$:
\be
f_0 = \int dk~k^2 \frac{\epsilon_0(k)}{f},~f_1 =
\int dk~k^2 \frac{f'^2}{f},~f_2
= \int dk~k^2 f,~f_4 = \int dk~k^4 f.\label{eq:mom}
\ee
For a general scale factor one obtains ($g=f_0+f_1/4$)
\bea
\rho_{g} &=& \frac{1}{2Ga^4} (f_4 +g +3\frac{a'}{a}f'_2 
-7\frac{a'^2}{a^2}f_2), \label{eq:rh1}\\
3 p_{g} &=& \frac{1}{2Ga^4} (7f_4 -5g +5\frac{a'}{a}f'_2 
-5\frac{a'^2}{a^2}f_2). \label{eq:pr1}
\eea
In the high-frequency approximation 
$3Ga^4p_g=Ga^4\rho_g= \int dk k^3 \sqrt{\epsilon_0}$, energy density and
pressure are positive. If low-frequency modes are present, their 
contribution can be negative, and also the equation of state can deviate
considerably from the high-frequency relation $p=\rho/3$.
If pressure and density of gravitational waves cannot be neglected 
compared to other forms of matter, the back-reaction on the scale 
factor must be taken into account. Taking a pure gravitational 
radiation universe, one has to solve the field equations 
\bea
6\frac{a'^{2}}{a^4}   &=& 16\pi G\rho_g, \label{eq:g0}\\
 \frac{a''}{a}+ \frac{a'^{2}}{a^2}
 &=& 4\pi G a^2(\rho_g -p_g), \label{eq:gik}
\eea
with $\rho_g,p_g$ taken from (\ref{eq:rh1}) and (\ref{eq:pr1}).
Note the further equations
\bea
 g'  &=& -f_4' + \frac{a''}{a}f_2', \label{eq:s1}\\
 f_2''&=& 2g +2\frac{a''}{a}f_2  -2f_4, \label{eq:s2}
\eea
which follow from differentiating $f_1, f_2$
and using the differential equation for $f$. The four
equations (\ref{eq:g0})-(\ref{eq:s2}) give the relation
\be
(2a'f_2 -af_2' )~(aa''-2 a'^2) = 0. \label{eq:prod}
\ee
Vanishing of the first factor leads to the Tolman radiation universe, 
vanishing of the second factor gives the scale factor of the de~Sitter
cosmos. It is easy to find the moments $f_2,f_4,g$ from the
corresponding differential equations. Self-consistency however requires,
that the moments found in this way must be compatible
with the expressions following directly from (\ref{eq:mom}), if the 
solution of the wave equation is inserted. Compatibility can be achieved 
indeed, it however requires singular infrared components for some spectral 
quantities. The general solution of (\ref{eq:al1}) for the radiation 
($s=0, a\sim \eta $) and 
de~Sitter ($s=1, a\sim 1/\eta$) cosmos is known to be
\be
f = 2npp^* +(l+im)p^2 +(l-im)p^{*2}, \label{eq:fall}
\ee
where $l,m,n$ are three functions of $k$, connected with $\epsilon_0$
by $\epsilon_0 = 4k^2(n^2-l^2-m^2)$,   
$p(x)= (1+is/x)e^{ix}$ is a complex function of $x= k\eta$.
Whereas $n(k)$ is not restricted, the spectral functions $l(k)$ and
$m(k)$ should be understood as generalized functions\cite{s} and 
have the form ($b$ is a constant)\footnote{for the Tolman cosmos, the 
condition is different for the de~Sitter case.}
\bea
l(k)&=& -\frac{b}{2}\frac{\delta''(k)}{k^2}
-2n_2\frac{\delta(k)}{k^2}, \\ m(k)&=& 0,
\eea
where $\delta(k)$ is the Dirac delta function and $n_2=\int dk k^2 n(k)$.
The spectral densities for the energy density and the pressure are 
then (again in the Tolman case)
\bea
a^4G\rho(k,\eta) &=& n(k)(k^2- \frac{7}{2\eta^2}) 
+ \delta(k)( \frac{7n_2}{\eta^2}- b),\\
3a^4Gp(k,\eta) &=& n(k)(k^2- \frac{5}{2\eta^2}) 
+ \delta(k)( \frac{5n_2}{\eta^2}- b).
\eea
These quantities show infrared singularities, 
part of the effective energy 
density and pressure resides in infrared ($k=0$) modes.  It is so far 
not clear whether the singularities result from using only a second-order 
approximation. 
The integrated quantities are finite, however.

\end{document}